\documentclass[aps,a4paper,12pt, twocolumns]{revtex4}
 \pdfoutput=1 
\usepackage{float}
\usepackage{fancyhdr}
\usepackage{color}
\usepackage{graphicx}
\usepackage{epsfig} 
\usepackage{amssymb}
\usepackage{amsmath,amsfonts,verbatim}
\usepackage{booktabs}
\usepackage{hyperref}
\usepackage{eufrak}
\usepackage{mathrsfs}
\usepackage[footnotesize,singlelinecheck=off,justification=raggedright]{caption}
\def\bra#1{\mathinner{\langle{#1}|}}
\def\ket#1{\mathinner{|{#1}\rangle}}

\def\Bra#1{\left<1>}
\def\g{{\rm g}}
\def\myPhi{\Theta}

\DeclareMathAlphabet{\mathbbmsl}{U}{bbm}{m}{sl}
{\catcode`\|=\active\gdef\Braket#1{\left<\mathcode`\|"8000\let|\bravert {#1}\right>}}
\def\bravert{\egroup\,\vrule\,\bgroup}

\def\Tr{\mathop{\mbox{\normalfont Tr}}\nolimits}

\def\chartheta#1#2{\vartheta\!\left[\begin{smallmatrix}#1\\#2\end{smallmatrix}\right]\!}
\def\hattheta#1#2{\widehat\vartheta\!\left[\begin{smallmatrix}#1\\#2\end{smallmatrix}\right]\!}
\begin{document}

\title{On the M\"obius transformation in 
the entanglement entropy of fermionic chains}

\author{Filiberto Ares\footnote{Corresponding author.}}
\email{ares@unizar.es}
\affiliation{Departamento de F\'{\i}sica Te\'orica, Universidad de Zaragoza,
50009 Zaragoza, Spain}
\author{Jos\'e G. Esteve}
 \email{esteve@unizar.es}
 \affiliation{Departamento de F\'{\i}sica Te\'orica, Universidad de Zaragoza,
50009 Zaragoza, Spain}
\affiliation{Instituto de Biocomputaci\'on y F\'{\i}sica de Sistemas
Complejos (BIFI), 50009 Zaragoza, Spain}
  \author{Fernando Falceto}
\email{falceto@unizar.es}
 \affiliation{Departamento de F\'{\i}sica Te\'orica, Universidad de Zaragoza,
50009 Zaragoza, Spain}
\affiliation{Instituto de Biocomputaci\'on y F\'{\i}sica de Sistemas
Complejos (BIFI), 50009 Zaragoza, Spain}
  \author{Amilcar R. de Queiroz}
\email{amilcarq@gmail.com}
 \affiliation{Departamento de F\'{\i}sica Te\'orica, Universidad de Zaragoza,
50009 Zaragoza, Spain}
 \affiliation{Instituto de Fisica, Universidade de Brasilia, Caixa Postal 04455, 70919-970, Bras\'{\i}lia, DF, Brazil}


\begin{abstract} 
There is an intimate relation between entanglement entropy and Riemann surfaces. 
This fact is explicitly noticed for the case of quadratic fermionic Hamiltonians 
with finite range couplings. After recollecting this fact, we make a comprehensive 
analysis of the action of the M\"obius transformations on the Riemann surface. 
We are then able to uncover the origin of some symmetries and dualities
of the entanglement entropy already noticed recently in the literature. 
These results give further support for the use of entanglement entropy to 
analyse phase transitions. 
\end{abstract}

\maketitle

\section{Introduction}

In the present work we make a comprehensive analysis of the symmetries 
and the dualities of the R\'enyi entanglement entropy for the ground state 
of quadratic, translational invariant fermionic Hamiltonians with finite 
range couplings in a one-dimensional chain. 
We here focus on the case of non-critical, that is, gapped Hamiltonians
with reflection and charge conjugation symmetry.

For such Hamiltonians, the entanglement entropy associated with a partial observation on an 
interval of contiguous sites is a functional of the determinant of the $2$-point correlation matrix. 
This matrix is of the block Toeplitz type \cite{Ares3}. It was shown \cite{Its,Its2} that by using 
Riemann-Hilbert problem techniques one can obtain the asymptotics of this determinant. 
Furthermore, in the process of computing the asymptotics of the determinant one is led 
to cast the moduli of Hamiltonians through their dispersion relations in terms of a compact Riemann surface. 
The branch points of the hyperelliptic curve which defines this Riemann surface depend 
on the coupling constants of the Hamiltonian. 

Our analysis of the symmetries and the dualities of the entanglement entropy is based on the study 
of the action of M\"obius transformations on the above Riemann surface, which moves the branch points and, therefore,
the coupling constants but leaves invariant
the asymptotic expansion of the determinant of 2-point correlations and, therefore, the entanglement entropy. 
We then obtain the following collection of remarkable results:

\begin{enumerate}
  
  \item We uncover the fact that the entanglement entropy only depends on the M\"obius transformation 
  invariants which are constructed with the branch points of the hyperelliptic curve 
  associated with the couplings of the model. This fact explains the ellipses of constant entropy 
  first noticed for the XY model in \cite{Franchini}.
  
  \item We analyse which M\"obius transformations map the original couplings to other
  physically allowed, i.e. to other that preserve the hermiticity of the Hamiltonian. 
  We find that these M\"obius transformations are inversions and 
  the 1+1 Lorentz group, $SO(1,1)$. In particular, we observe that under $SO(1,1)$ 
  the couplings change such that the dispersion relation behaves as a homogeneous field 
  with scaling dimension equal to the maximum range $L$ of the coupling. 
  
  \item We also note that the asymptotic expansion of the determinant is also invariant
  under certain permutations of the branch points which induce a modular transformation of
  the non contractible cycles of our Riemann surface. 
  This fact, combined with the previous invariance under M\"obius transformations, can be useful
  to find and analyse symmetries and dualities between different Hamiltonians in terms of entanglement entropy.
  As an example, we apply this idea to the XY spin chain. In particular, the duality noticed 
  in \cite{Franchini} is discussed in terms of a Dehn twist of the underlying Riemann 
  surface which in this XY case is a torus. Recall that two distinct Dehn twists, 
  which are two special transformations of the moduli of the torus, 
  generate the whole modular group in that case. 
  The Dehn twists play an important role in string theory, specially in the analysis 
  of its many dualities. See for instance \cite{Blumenhagen, AlvarezGaume}.
  Another suggestive work is \cite{Ercolessi1} where the modular 
  group is also used to understand the dualities 
  of the XYZ spin chain. 
  
  \item This framework also allows to extend the relation between 
  the entanglement entropy of two Kramers-Wannier dual Ising models and its 
  corresponding XX model, discussed by Igloi and Juhasz in \cite{Igloi}, 
  to general XY models.

\end{enumerate}

We organize this paper as follows. In section \ref{frh} we first present the most general 
quadratic fermionic Hamiltonian with long range couplings, reflection and charge conjugation symmetry,
in a chain with $N$ sites. We next obtain the corresponding dispersion relation and then 
write a formula for the R\'enyi entanglement entropy in terms of the determinant of the $2$-point correlation matrix
for the ground state. Following the notorious works \cite{Its, Its2}, we take the asymptotics 
of this determinant (\ref{determinant}) by constructing a compact Riemann surface 
from the coupling constants of the model. It is this Riemann surface that will allow us to analyse the symmetries 
and dualities of the entanglement entropy based on its behaviour under the M\"obius transformations. 
In section \ref{mt-section} we make a comprehensive analysis of the M\"obius transformations on 
the Riemann surface associated with the moduli of coupling constants. We obtain that the entanglement entropy is left invariant, 
which allows us to obtain entropy preserving dualities between Hamiltonians. 
Furthermore, the only M\"obius transformations
that preserve the properties of the Riemann surface imposed by the couplings
are inversion and the 1+1 Lorentz group. A beautiful outcome of this result is
that under the latter the dispersion relation itself behaves as a homogeneous 
field with scaling dimension associated with the maximum range $L$ of the couplings. 
In section \ref{xymodel-section} we apply our general results of previous section 
to the XY model. We are then able to uncover the geometrical origin of some dualities
already noticed in \cite{Franchini, Igloi} and to generalize them. Finally, we conclude in section \ref{conclusion-section} 
by summarizing our findings and discussing some prospects for the future. 

This paper also contains two appendices. In appendix \ref{det-invariance-app}, we show that the 
determinant of our block Toeplitz matrix is invariant under permutation of the branch points defining the Riemann surface. 
In appendix \ref{vilenkin-app}, we recollect the important facts concerning the representation of $SL(2,\mathbb{C})$ 
on the space of homogeneous polynomials of two complex variables. This plays a crucial role in the understanding 
how the Hamiltonian, and in particular its couplings constants, change under M\"obius transformations.

\section{Finite range Hamiltonians, compact Riemann surfaces and entanglement entropy}\label{frh}

Let us consider a $N$-site chain of size $\ell$ with $N$ spinless 
fermions described by a quadratic, translational invariant Hamiltonian 
with finite range couplings ($L<N/2$)
\begin{equation}\label{hamiltonian} 
 H=\frac12
\sum_{n=1}^N \sum_{l=-L}^L ~\left( 2 A_l a_n^\dagger a_{n+l}+B_l a_n^\dagger a_{n+l}^\dagger
- B_l a_n a_{n+l} \right).
\end{equation}
Here $a_n$ and $a_n^\dagger$ represent the annihilation and creation operators at the site $n$. 
The only non-vanishing anticommutation relations are
 $$\{a_n^\dagger, a_m\}=\delta_{nm}.$$
We assume periodic boundary conditions $a_{n+N}=a_n$ in (\ref{hamiltonian}).

We take $A_l$ and $B_l$ real, and in order that $H$ is Hermitian, 
the hopping couplings must be symmetric $A_{-l}={A}_{l}$. 
In addition, without loss of generality, we may take $B_l=-B_{-l}$. 
For a comprehensive analysis of more general choices of 
couplings $A_l$ and $B_l$ see \cite{Ares3}.

Proceeding like in \cite{Ares3} we introduce
 \begin{equation}
	 \label{Fk-def-1}
   F_k=\sum_{l=-L}^L A_l e^{i\theta_k l}, \qquad \theta_k=\frac{2\pi k}{N}, 
 \end{equation}
and
 \begin{equation}
	 \label{Gk-def-1}
   G_k=\sum_{l=-L}^L B_l e^{i\theta_k l}.
 \end{equation}
Note that in our case, $F_k$ is real and $F_{-k}=F_k$ while $G_k$ is imaginary and $G_{-k}=-G_k$.

The Hamiltonian can now be written in diagonal form by means 
of anticommuting annihilation, creation operators $d_k, d^\dagger_k$,
the Bogoliubov transform of the Fourier modes,
 $$H={\cal E}+\sum_{k=0}^{N-1} \Lambda_k
\left(d_k^\dagger d_k-\frac12\right),$$
where 
\begin{equation}
	\label{Evar-def-1}
  {\cal E}=\frac12\sum_{k=0}^{N-1}F_k.
\end{equation}
The dispersion relation reads
\begin{equation}
  \label{Dis-Rel-1}
  \Lambda_k=\sqrt{(F_{k}+G_k)(F_k-G_k)},
\end{equation} 
which takes non-negative values. Hence the ground state of the theory $\ket{\rm GS}$ 
is the vacuum of the Fock space for the $d_k$ modes, i.e. $d_k\ket{\rm GS}=0$, $\forall k$.

In this paper we focus our study in the R\'enyi entanglement entropy for the ground state of Hamiltonian (\ref{hamiltonian}).

Given a subset $X$ of contiguous sites of the fermionic chain, with complementary set $Y$ we 
have the corresponding factorization of the Hilbert space ${\cal H}={\cal H}_X\otimes{\cal H}_Y$. 
If the system is in the pure state $\ket{\mathrm{GS}}$, the reduced density matrix for $X$ is obtained by 
taking the partial trace with respect to ${\cal H}_Y$, $\rho_X=\Tr_Y\ket{\mathrm{GS}}\bra{\mathrm{GS}}$. 
The R\'enyi entanglement entropy of $X$ is
\begin{equation}
  \label{Renyi-ent-1}
  S_\alpha(X)=\frac1{1-\alpha}\log \Tr(\rho_X^\alpha).
\end{equation}

As it is well known \cite{Latorre, Peschel, Jin, Balachandran1, Balachandran2}, 
due to the fact that the $n$-point vacuum expectation value satisfies the Wick decomposition property, 
it is possible to derive the reduced density matrix from the two point 
correlation function. Namely, for any pair of sites $n,m\in X$, we introduce the correlation matrix 
 $$(V_X)_{nm}=2\left< \left( \begin{array}{c} a_n \\ a_n^\dagger \end{array}\right)
 (a_m^\dagger, a_m)\right>-\delta_{nm}\; I =
 \left(\begin{array}{cc} \delta_{nm}-2\left<a_m^\dagger a_n\right> & 2\left<a_n a_m\right> \\ 
 2\left<a_n^\dagger a_m^\dagger\right> & 2\left<a_n^\dagger a_m\right>-\delta_{nm} \end{array}\right).$$

Following \cite{Latorre, Peschel, Jin} one can show that the R\'enyi entanglement entropy reads
 \begin{equation}\label{entcorrel}
 S_\alpha(X)=\frac{1}{2(1-\alpha)}\Tr\log\left[\left(\frac{\mathbb{I}+V_X}{2}\right)^\alpha
 +\left(\frac{\mathbb I-V_X}{2}\right)^\alpha\right],
 \end{equation}
where $\mathbb I$ is the $2|X|\times 2|X|$ identity matrix and $|X|$ denotes the size of $X$.

\begin{figure}[h]
  \centering
    \resizebox{12cm}{4cm}{\includegraphics{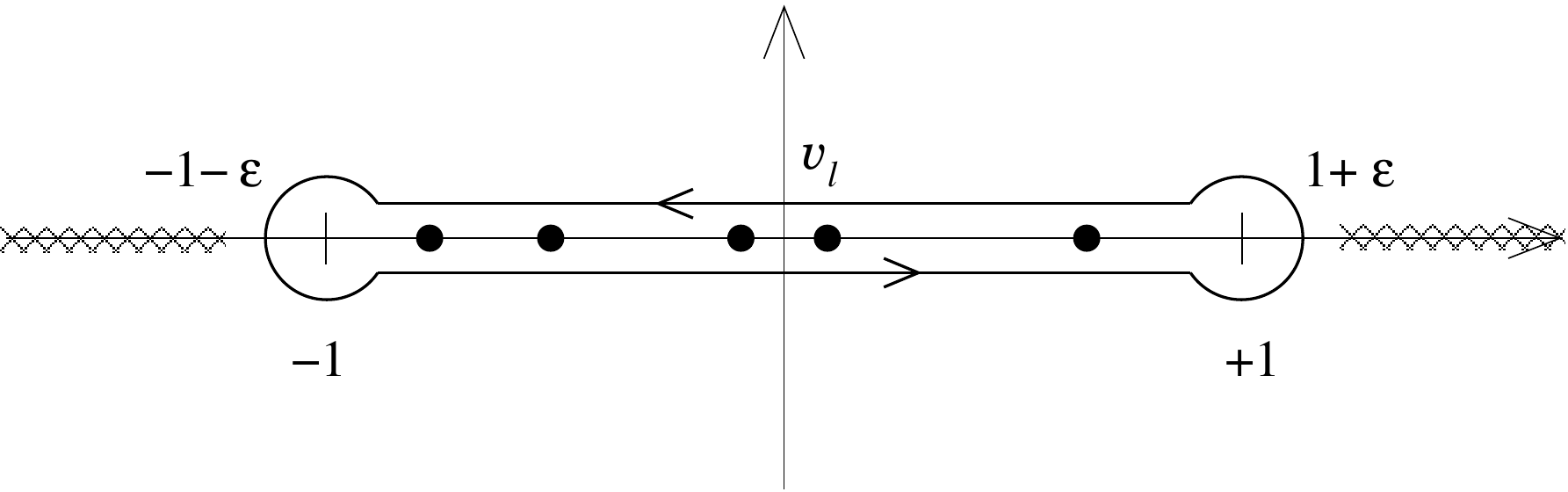}} 
    \caption{Contour of integration, cuts and poles for the computation of 
$S_\alpha(X)$. The cuts for the function $f_\alpha$ extend to $\pm\infty$.}
  \label{contorno0}
   \end{figure}

The Cauchy's residue theorem allows us to implement (\ref{entcorrel}) as
 \begin{equation}\label{entintegr}
 S_\alpha(X)=\lim_{\varepsilon\to0^+}\frac{1}{4\pi i}\oint_{\mathcal{C}}
 f_\alpha(1+\varepsilon, \lambda)\frac{\mathrm{d}
\log D_X(\lambda)}
{\mathrm{d}\lambda}
 \mathrm{d}\lambda,
\end{equation}
 where $D_X(\lambda)=\det(\lambda {\mathbb I}-V_X)$,
\begin{equation}\label{efealfa}
f_\alpha(x, y)=\frac{1}{1-\alpha}\log\left[\left(\frac{x+y}{2}\right)^\alpha+\left(\frac{x-y}{2}\right)^\alpha\right],
\end{equation}
 and $\mathcal{C}$ is the contour depicted in Fig. \ref{contorno0} which 
 surrounds the eigenvalues $v_l$ of $V_X$, all of them lying in the real interval $[-1,1]$. 

Applying to our case the results in \cite{Ares3} one can show that
$$(V_X)_{nm}=\frac{1}{N}\sum_{k=0}^{N-1} M_k e^{i\theta_k (n-m)},$$
with
$$M_k
=\frac{1}{\Lambda_k}
 \left(\begin{array}{cc} F_{k} & G_k \\ -G_k & -F_{k} \end{array}\right).$$
We now consider the thermodynamic limit, $N\to\infty$, $\ell\to\infty$ with $N/\ell$ fixed.
In this limit the previous $N$-tuples, like $\Lambda_k=\Lambda(\theta_k), F_k=F(\theta_k), G_k=G(\theta_k)$ 
and the others, with $\theta_k=2\pi k/N$, become $2\pi$-periodic 
functions of the continuous variable $\theta$, that is, $\Lambda(\theta), F(\theta), G(\theta),\dots$. 

In order to obtain the asymptotic behaviour of the entanglement entropy  (\ref{entintegr}), we must
compute $D_X(\lambda)\equiv \det(\lambda {\mathbb I}-V_X)$ when the size $|X|$ of $X$ is large.

This task was solved in \cite{Its,Its2} by reducing it to a
particular Wiener-Hopf factorization problem. To describe the result we introduce the (complex) Laurent polynomials 
\begin{equation}\label{laurent}
\myPhi(z)=\sum_{l=-L}^LA_l z^l,\quad \Xi(z)=\sum_{l=-L}^LB_l z^l,
\end{equation}
that are related to $F$ and $G$, from (\ref{Fk-def-1}) and (\ref{Gk-def-1}), 
by $F(\theta)=\myPhi({\rm e}^{i\theta})$ and $G(\theta)=\Xi({\rm e}^{i\theta})$.
In terms of (\ref{laurent}) we extend the symbol $M(\theta)$ to the complex 
plane
$${\cal M}(z)=  \frac1{\sqrt{\myPhi^2(z)-\Xi^2(z)}}\left(\begin{array}{cc} 
\myPhi(z) & \Xi(z) \\ -\Xi(z) & -\myPhi(z) \end{array}\right)=
U\left(\begin{array}{cc} 0 & g(z) \\ g(z)^{-1} & 0 \end{array}\right)U^{-1},$$
with
$$U=\frac1{\sqrt{2}}\left(\begin{array}{cc} 1 & 1 \\ -1 & 1 \end{array}\right)$$ and
$$g(z)=\sqrt{\frac{\myPhi(z)+\Xi(z)}{\myPhi(z)-\Xi(z)}}.$$

Now, as it is shown in \cite{Widom,Its,Its2},
the asymptotic behaviour of the logarithmic derivative of 
$D_X(\lambda)$ verifies
\begin{eqnarray}\label{wiener-hopf}
\frac{\rm d}{{\rm d} \lambda}
\ln D_X(\lambda)
&=&
\frac{2\lambda}{\lambda^2-1}|X|+\cr\cr&&
\hskip -5mm+\frac1{2\pi}\int_{|z|=1}\Tr\left[\left(u_+'(z)u_+^{-1}(z)+
v_+^{-1}(z)v_+'(z)\right){\cal M}^{-1}(z)\right]{\rm d}z
+\dots
\end{eqnarray}
where the prime denotes the derivative with respect to $z$ and 
$u_+, v_+$ are the solution to the following {\bf Wiener-Hopf} factorization
problem: 

i) ${\cal M}(z)=u_+(z)u_-(z)=v_-(z)v_+(z)$,

ii) with $u_-^{\pm1}(z), v_-^{\pm1}(z)$ analytic outside the unit circle and 
$u^{\pm1}_+(z), v^{\pm1}_+(z)$ analytic inside.  
\vskip 2mm

This problem has been solved in \cite{Its, Its2} and the results,
expressed in terms of theta functions associated to the
Riemann surface determined by $g(z)$, are described in the following.

First, consider the analytic structure of $g(z)$. 
Actually it is a bivalued function in the complex plane, but
it is a single valued meromorphic function in the Riemann surface
determined by the complex curve 
\begin{equation}\label{Riemann}
w^2=P(z)\equiv z^{2L}(\myPhi(z)+\Xi(z))(\myPhi(z)-\Xi(z)),\quad w,z\in
\overline{\mathbb C},
\end{equation}
where $\overline{\mathbb C}$ denotes the  Riemann sphere.

Here we shall assume that the polynomial $P(z)$ has $4L$ different simple roots
and therefore (\ref{Riemann}) defines a two-sheet Riemann surface of genus $\g=2L-1$.
On the other hand, since the real coupling constants of (\ref{hamiltonian}) 
satisfy $A_l=A_{-l}$ and $B_{-l}=-B_l$, we have
$$\myPhi(z^{-1})=\myPhi(z),\quad \overline{\myPhi(z)}=\myPhi(\overline z),
\quad \Xi(z^{-1})=-\Xi(z), \quad\overline{\Xi(z)}=\Xi(\overline z).$$
Therefore if $z_j$ is a root of $P(z)$ then $\overline z_j$ and $z_j^{-1}$ are roots as well.
Actually the roots of $P(z)$ coincide with the zeros and poles of the rational function
$g^2(z)$ and the previous relation specializes as follows:
if $z_j$ is a zero (pole) of $g^2(z)$ then $\overline z_j$ is also a zero (pole)
while $z_j^{-1}$ is a pole (zero). 

We fix an order in the roots of $P(z)$ with the only requierement that the first half 
of them is inside the unit circle and the other half outside, i.e.
$|z_j|<1, j=1,\dots,2L$ and $|z_j|>1, j=2L+1,\dots,4L$
(notice that if all roots are simple, then necessarily $|z_j|\not=1$).

Equipped with the previous data we fix the branch cuts
of $g(z)$, which are $2L$ non intersecting curves
$\Sigma_\rho,\ \rho=0,\dots,\g$ that join $z_{2\rho+1}$ and $z_{2\rho+2}$. 
Notice that it is always possible to choose them such 
that they do not cross the unit circle.

\begin{figure}[h]
  \centering
     \resizebox{16cm}{7cm}{\includegraphics{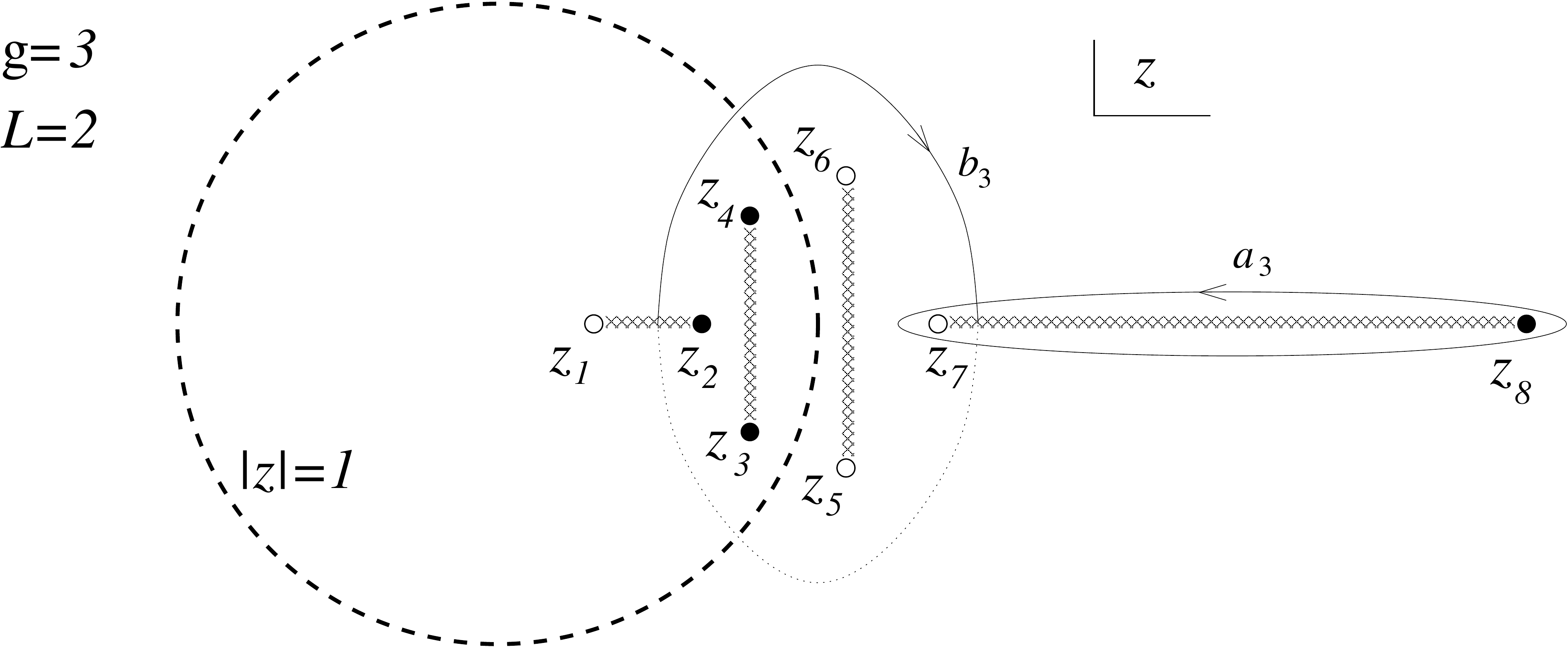}}
    \caption{Possible arrangement of the branch points and cuts of $w=\sqrt{P(z)}$ for
genus $\g=3$ ($L=2$).
    Note that we must have $z_1=z_8^{-1}$, $z_2= z_7^{-1}$ ,  $z_3=\overline{z}_4=\overline z_5^{-1}=z_6^{-1}$. 
    The branch points in black ($\bullet$) are zeros of $g^2(z)$ ($\epsilon_j=1$) 
while those in white ($\circ$) are poles ($\epsilon_j=-1$). Oriented curves $a_3$ and $b_3$ are
  two of the basic cycles.}
  \label{cuts}
   \end{figure}

Associated to these cuts we have a canonical homology basis 
of cycles in the Riemann surface $a_r, b_r,\ r=1,\dots,\g$
where $a_r$, in the upper Riemann sheet, surrounds $\Sigma_r$ anticlockwise
and the dual cycle $b_r$ surrounds the branch points 
$z_2, z_3,\cdots,z_{2r+1}$ clockwise. 
In Fig. \ref{cuts} we 
depict a possible arrangement of the branch points and cuts 
for $\mathrm{\g}=3$ ($L=2$) as well as cycles $a_3$ and $b_3$.

   The canonical basis of holomorphic forms 
$$\label{canonical-forms}\mathrm{d}\omega_r=\frac{\varphi_r(z)}{\sqrt{P(z)}}\,{\rm d}z,$$
with $\varphi_r(z)$ a polynomial of degree smaller than $\g$, 
is chosen so that $\int_{a_r} {\rm d}\omega_{r'}=\delta_{rr'}$.
The $\g\times\g$ symmetric matrix of periods $\Pi=(\Pi_{rr'})$ 
is given by
$$\label{periodic-matrix}\Pi_{rr'}=\int_{b_r} {\rm d}\omega_{r'}.$$
Associated with this matrix of periods we can now define the Riemann theta-function with characteristics $\vec p,\vec q\in {\mathbb R}^\g$, 
$\chartheta{\vec p}{\vec q}:{\mathbb C}^\g \rightarrow{\mathbb C}$,
given by
$$
\chartheta{\vec p}{\vec q}(\vec s)
\equiv\chartheta{\vec p}{\vec q}(\vec s\,|\,\Pi)=
\sum_{\vec{n}\in \mathbb{Z}^{^\g}}~e^{\pi i(\vec n+\vec p)\Pi\cdot(\vec n+\vec p)+
2\pi i(\vec s+\vec q)\cdot(\vec n+\vec p)},$$
and its {\it normalized} version
$$
\hattheta{\vec p}{\vec q}(\vec s)\equiv\hattheta{\vec p}{\vec q}(\vec s\,|\,\Pi)=\frac{\chartheta{\vec p}{\vec q}(\vec s\,|\,\Pi)}
{\chartheta{\vec p}{\vec q}(0\,|\,\Pi)}.
$$

Finally, one has the following expression for the asymptotic value of 
the determinant (adapted from Proposition 1 of \cite{Its2}):
\begin{equation}\label{determinant}
\log{D_X(\lambda)}
=
{|X|\log(\lambda^2-1)}+\log\left(
\hattheta{\vec\mu}{\vec\nu}(\beta(\lambda)\vec e)
\ \hattheta{\vec\mu}{\vec\nu}(-\beta(\lambda)\vec e)
\right)+\cdots,
\end{equation}
where the dots represent terms that vanish in the large $|X|$ limit.
The argument of the theta function contains
\begin{equation}\label{beta}
\beta(\lambda)=\frac1{2\pi i}\log\frac{\lambda+1}{\lambda-1},
\end{equation}
and $\vec{e}\in\mathbb{Z}^{\g}$, a $\g$ dimensional vector whose first $L-1$ entries are 0 and
the last $L$ are 1.
The characteristics of the theta function are half integer vectors 
$\vec\mu,\vec\nu\in({\mathbb Z/2})^\g$ 
which are determined as follows:
first assign an index $\epsilon_j$ to every root $z_j$ of $P(z)$,
so that $\epsilon_j=1$ ($\epsilon_j=-1$) if $z_j$ is a zero (pole) of $g^2(z)$;
then the vectors are
\begin{eqnarray}\label{characteristics}
\mu_r&=&\frac14(\epsilon_{2r+1}+\epsilon_{2r+2}),\nonumber\\
\nu_r&=&\frac14\sum_{j=2}^{2r+1}\epsilon_{j},\qquad r=1,\dots,\g.
\end{eqnarray}
Note that the indices must satisfy $\epsilon_j=-\epsilon_{j'}$ whenever
$z_j=z_{j'}^{-1}$ and $\epsilon_j=\epsilon_{j'}$ if
$z_j=\overline z_{j'}$.

In order to give an explicit expression for the determinant we have 
fixed an order of the roots with the only requirement that the $2L$ 
first ones are inside the unit circle and the last ones outside. 
Of course, for consistency (\ref{determinant}) should not depend on the chosen order. 
Actually it is an instructive exercise to show that the value 
of the determinant (in the thermodynamic limit) is in fact invariant under 
the transposition of two roots, provided they sit at the same side of 
the unit circle. In appendix \ref{det-invariance-app} we outline a proof of this fact.

\section{M\"obius transformation}\label{mt-section}

If we now perform a holomorphic bijective transformation in the Riemann sphere, $z'=f(z)$,  
we move the branch points and cuts, thus modifying the holomorphic forms. 
But it is clear that the matrix of periods is unchanged. 
Therefore the theta function does not change
and the entropy 
derived from (\ref{determinant}) is left invariant.

It is a mathematical fact that the only holomorphic one-to-one 
maps of the Riemann sphere into itself are the M\"obius transformations 
$$z'=\frac{az+b}{cz+d},\qquad \left(\begin{array}{cc}a&b\\c&d\end{array}\right)\in SL(2,{\mathbb C}).$$
They act, see Appendix \ref{vilenkin-app}, on the Laurent polynomial $\myPhi$ by 
\begin{equation}\label{moebius-laurent}
\myPhi'(z')=(az+b)^{-L}(dz^{-1}+c)^{-L}\myPhi(z),
\end{equation}
which is again a Laurent polynomial with monomials of degree
between $L$ and $-L$. Hence, the M\"obius transformation can be seen
as a change of the couplings from $A_l$ to $A'_l$. 
In exactly the same way we transform $\Xi$ with coefficients
$B_l$ to a new $\Xi'$ with coefficients $B'_l$.

Notice that if we use the new Laurent polynomials to get the new rational function 
$$g'^2=\frac{\myPhi'+\Xi'}{\myPhi'-\Xi'},$$
we have $g'^2(z')=g^2(z)$.

But this is not the end of the story because as we mentioned in the
previous section, the roots of
$P(z)$ (\ref{Riemann}) satisfy certain properties, namely they 
come in quartets $z_j$, $\overline z_j$, $z_j^{-1}$ and $\overline z_j^{-1}$.
In order to preserve this property, we must require that the M\"obius 
transformation commute with the complex conjugation and inversion.

The commutation with conjugation restricts $SL(2,{\mathbb C})$ to
the semidirect product $SL(2,{\mathbb R})\rtimes \{I,i\sigma^x\}$.
These are the transformations that preserve the real line.
Note that the second factor in the semidirect product
is related to the inversion $z'=1/z$ and maps the upper 
half plane into the lower one while the first factor contains the transformations that map the upper half plane to itself.

If we further impose that the allowed transformations must  commute with
the inversion, we are finally left with the group generated by the 
reflection $z'=1/z$  and the 1+1 Lorentz group $SO(1,1)$
whose elements act on $z$ by 
\begin{equation}\label{mymoebius}
z'=\frac{z \cosh\zeta +\sinh\zeta }{z \sinh\zeta +\cosh\zeta },\qquad\zeta\in{\mathbb R}.
\end{equation}
This is precisely the subgroup of the M\"obius tranformations
that preserve the unit circle and the real line. 
Its connected component maps the upper half plane and the unit disc 
into themselves. 

We unraveled the symmetry by considering the final expression for the 
asymptotic determinant (\ref{determinant}), but one can trace back its 
origin to the Wiener-Hopf factorization problem described in 
the previous section. In fact, assume we perform a M\"obius transformation 
that preserve the unit circle mapping its exterior into its exterior
and the interior into the interior. Then, 
the property of being analytic outside (inside) the circle is 
preserved which implies that the solution to the Wiener-Hopf
factorization for ${\cal M}(z)$ is transformed into that for ${\cal M'}(z)$.
Inserting it into (\ref{wiener-hopf}) one immediately sees
that the logarithmic derivative of $D_X(\lambda)$ 
and hence the entanglement entropy,  
are unchanged, at least in the asymptotic limit.
Actually, a similar reasoning shows that the determinant is 
invariant for any M\"obius transformation provided it keeps inside (outside)
the unit circle the branching points that were
originally inside (outside).

As for the physical interpretation of the transformations, 
the inversion corresponds to reverting the orientation in 
the chain $n\leftrightarrow N-n$ which, clearly, does not 
affect the entropy of the system.
The coupling constants transform in a very simple way, namely, $A'_l=A_{l}$,
$B'_l=-B_{l}$. The symmetries in the Lorentz group, however, have a rather non trivial 
implementation. In Appendix B we shall discuss in full generality how the coupling 
constants of the Hamiltonian behave under such Lorentz transformations. 

On the other hand, it is interesting to observe that these particular M\"obius 
transformations leave invariant not only the von Neumann entropy but also the 
R\'enyi entanglement entropy (\ref{Renyi-ent-1}) for any value of $\alpha$.
Actually, all the spectral properties of the two-point correlation function
are preserved, at least in the large $|X|$ limit.

A very important observation is that all the previous does not apply for the critical,
gapless theories, i.e. when couples of roots of $P(z)$ coincide at the unit circle. This important case will be 
studied separately in a future publication \cite{Ares4}. 

With respect to the dynamical aspects of these transformations, we may say that they
are not a symmetry of the Hamiltonian. In the thermodynamic limit 
they act as a rescaling of the spectrum. Actually, 
as the unit circle is mapped into itself,
we may view the transformation as a change in the momentum of the modes 
together with a rescaling of its energy. 
More concretely, adopting again the active view point we have
$$\Lambda'(\theta')=\left(\frac{\partial\theta'}{\partial\theta}\right)^L
\Lambda(\theta),$$ where $\theta'$ is the image of $\theta$ under
the M\"obius transformation, 
$${\rm e}^{i\theta'}=\frac{{\rm e}^{i\theta}\cosh 
\zeta+\sinh\zeta}{{\rm e}^{i\theta}\sinh \zeta+\cosh\zeta},$$
and, therefore,
$$\frac{\partial\theta'}{\partial\theta}
=\frac1{\sinh2\zeta\cos\theta+\cosh2\zeta}.$$
Interestingly enough under the above transformations
the dispersion relation $\Lambda(\theta)$ behaves as a homogeneous
field of dimension $L$. Moreover, the dimension is directly associated with the range of the coupling.
We recall that such a transformation actually corresponds to a change 
of the coupling constants of the theory.

\begin{figure}[H]
  \centering
     \resizebox{8.5cm}{8.5cm}{\includegraphics{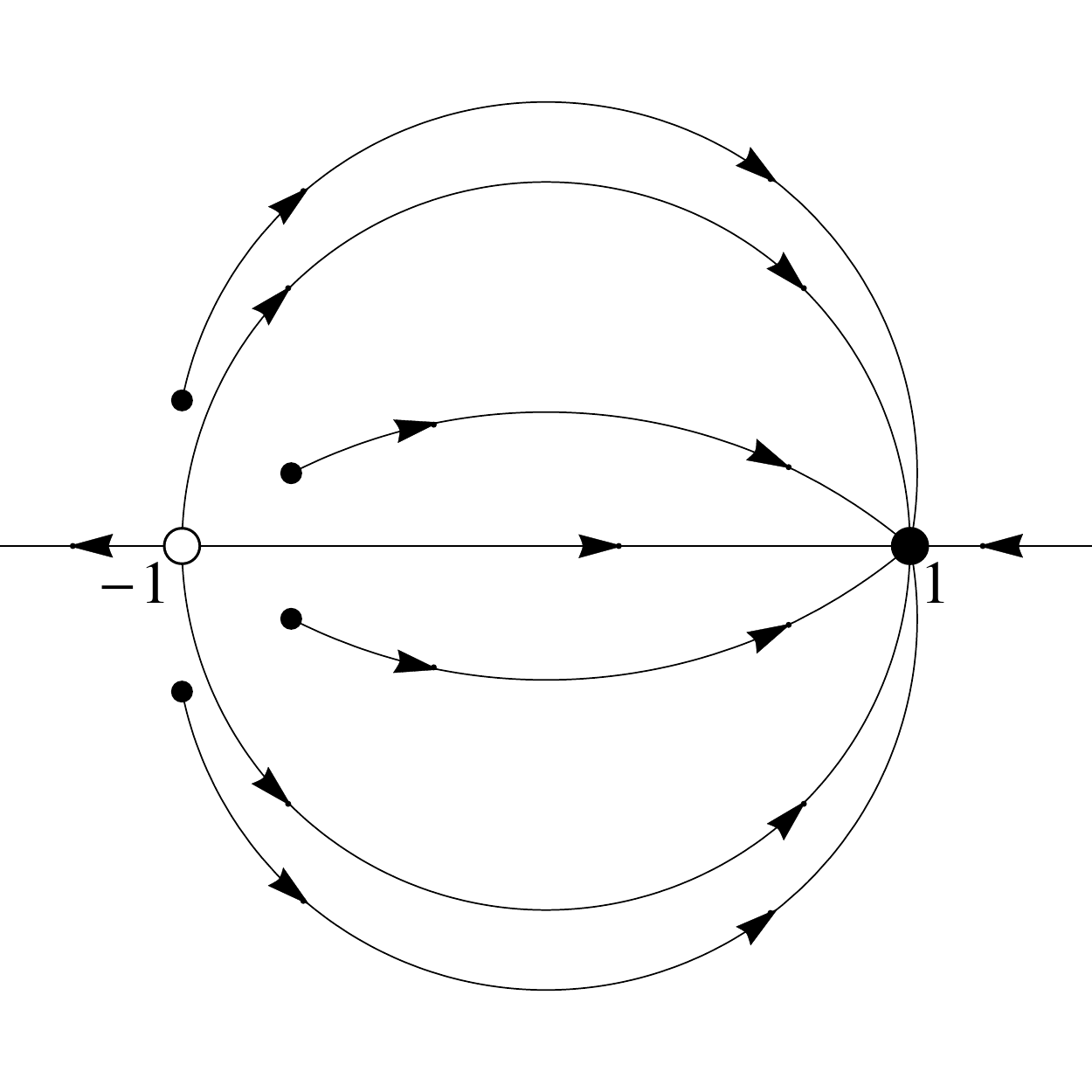}}
    \caption{Flow of $SO(1,1)$ in $\mathbb{C}$. Note that it preserves
    the unit circle and the real line and maps the unit disc and the upper/lower half plane into
    themeselves. The points $z=\pm1$ are the fixed points
    of the transformation; $+1$ is stable and $-1$ unstable.}
  \label{flow}
   \end{figure}

The action of $SO(1,1)$ in $\overline{\mathbb C}$ has two fixed points in $z=\pm1$. 
In particular, $1$ is stable and $-1$ is unstable,
so all the flow lines of the transformation depart from the first one and they join
in the second one as we sketch in Fig. \ref{flow}. 

If one considers the induced flow in the space of Hamiltonians
the picture is richer. There exist $2L+1$ fixed points
whose associated complex curves are
$$w^2=P(z)\equiv(z-1)^{4L-2j} (z+1)^{2j}, \quad j=0,\dots, 2L.$$
The only stable one corresponds to $j=0$. 
The other fixed points have a $j$-dimensional unstable manifold.
Of course, all these Hamiltonians correspond to critical theories.

\vskip 2mm
Summarizing, we have found a transformation of the coupling constants of a
free fermionic chain with finite range coupling such that the spectral 
properties of the two-point correlation function
are left invariant in the thermodynamic limit,
provided the theory has a mass gap. 
As a byproduct we show that the asymptotic behaviour of the R\'enyi 
entanglement entropy is unchanged, whenever it is finite. 
On the other hand, the modes in the one particle energy spectrum are
rescaled with a dimension given by the range of the couplings
of the theory.
The connected component of the symmetry group is composed by the M\"obius 
transformations associated to the 1+1 Lorentz subgroup $SO(1,1)$.  

Here we have only focused in a single connected interval. 
It is natural to take several disjoint intervals 
\cite{Furukawa,Calabrese2,Calabrese3,Facchi,Fagotti1}. 
In this case the situation is more involved because
the correlation matrix of the subsystem is no longer of the block Toeplitz type.
Nevertheless, our preliminary numeric computations suggest that the formulae
obtained in \cite{AEF} are still valid in this case. This would imply that 
the M\"obius symmetry also holds when $X$ is made of several disjoint 
intervals.

\section{The XY spin chain}\label{xymodel-section}

Our first application of the transformations introduced in the previous section is to
understand the different dualities and invariances of the entanglement entropy in the XY model. 
This has been studied in \cite{Franchini}. Here we rederive their results and find some new ones
using the ideas of the previous section.

The Hamiltonian reads
$$H_\mathrm{XY}=\frac{1}{2}\sum_{n=1}^N\left[(1+\gamma)\sigma_{n}^x\sigma_{n+1}^x+
 (1-\gamma)\sigma_n^y\sigma_{n+1}^y-
h\sigma_n^z\right].$$
The coupling constants $h$ and $\gamma$ are assumed to be real and positive. 

A Jordan-Wigner transformation allows us to write this
Hamiltonian in terms of fermionic operators, namely
 \begin{equation}\label{dmhamiltonian}
  H_{\mathrm{XY}}=\sum_{n=1}^N 
\left[a_n^\dagger a_{n+1}+a_{n+1}^\dagger a_{n}
  +\gamma (a_n^\dagger a_{n+1}^\dagger-a_na_{n+1})-h a_n^\dagger a_n\right]+\frac{Nh}{2},
 \end{equation}
which is a particular case of the Hamiltonian 
of (\ref{hamiltonian}) with nearest neighbours coupling, $L=1$. We are now set to apply the previous results.

The Laurent polynomials are in this case
 $$\myPhi(z)=z-h+ z^{-1},\quad
\Xi(z)=\gamma (z-z^{-1}),$$
and the dispersion relation is
\begin{equation}\label{dispersion}
\Lambda(\theta)=\sqrt{(h-2\cos \theta)^2+4\gamma^2\sin^2\theta}.
\end{equation}
The theory is gapless for $h=2$ (Ising universality class) 
or $\gamma=0, h<2$ (XX universality class). 

The zeros of $g^2(z)=(\myPhi(z)+\Xi(z))/(\myPhi(z)-\Xi(z))$ 
are given by 
\begin{equation}\label{roots}
z_\pm=\frac{h/2\pm\sqrt{(h/2)^2+\gamma^2-1}}{1+\gamma},
\end{equation}
and its poles are the inverses $z_{\pm}^{-1}$.
Note that the critical theories correspond to the value of the couplings
for which some of the zeros are in the complex unit circle. 
Here we do not discuss the case of critical theories, which deserves a special treatment \cite{Ares4}. 
Our focus here is on the non-critical theories, 
its symmetries, dualities and other properties.

If we now apply the transformations (\ref{mymoebius}) the couplings of the theory change as
\begin{eqnarray}\label{ellipse}
\gamma'&=&\frac{\gamma}{(h/2)\sinh 2\zeta+\cosh 2\zeta},\\
h'/2&=&\frac{(h/2)\cosh 2\zeta+\sinh 2\zeta}{(h/2)\sinh 2\zeta+\cosh 2\zeta},
\end{eqnarray}
while the zeros change as
$$z_{\pm}'=\frac{z_{\pm}\cosh\zeta+\sinh\zeta}{z_{\pm}\sinh\zeta+\cosh\zeta},$$
and similarly for the poles.

Actually, as it was discussed in the previous section, the entropy 
derived from (\ref{entintegr}) and (\ref{determinant})
is invariant under any M\"obius transformation of the Riemann surface.
Therefore it only depends on the zeros and poles of $g^2(z)$ 
through the  M\"obius invariants. For the XY model in which we have 
just four such points (two zeros $z_+, z_-$ and two poles 
$z_+^{-1},z_-^{-1}$) the only invariants are functions of the cross ratio
\begin{equation}\label{crossratio}
x\equiv(z_+,z_-;z_+^{-1},z_-^{-1})=\frac{(z_+-z_+^{-1})(z_--z_-^{-1})}{(z_+-z_-^{-1})(z_--z_+^{-1})}.
\end{equation}
If we use (\ref{roots}), this cross ratio can be written in terms of the couplings as
\begin{equation}\label{equx}
x=\frac{1-(h/2)^2}{\gamma^2}.
\end{equation}
The explicit form of the von Neumann entanglement entropy for non-critical 
XY model was computed in \cite{Its} and it was rewritten
in \cite{Franchini, Ares3} in terms of the parameter $x$. See also \cite{Peschel2, Franchini2}. 
We express the von Neumann entropy in three distinct regions, namely 1a, 1b and 2:
\begin{itemize}
\item Region 1a: $0<x<1$.
\begin{equation}\label{region1a}
S_1=\frac16\left[\log\left(\frac{\ 1-{x\;\ }}{16\sqrt{\ x\;\ }}\right)
+\frac{\ 2(1+{x})\;\ }\pi I(\sqrt{\ 1-{x\;\ }})I({\sqrt{\ x\;\ }})\right]+\log 2.
\end{equation}

\item Region 1b: $x>1$.
\begin{equation}\label{region1b}
S_1=\frac16\left[\log\left(\frac{1-{x{^{^{-1}}}}}{16\sqrt{x{^{^{-1}}}}}\right)
+\frac{2(1+{x{^{^{-1}}}})}\pi I(\sqrt{1-{x{^{^{-1}}}}})I({\sqrt {x{^{^{-1}}}}})\right]+\log 2.
\end{equation}

\item Region 2: ${x}<0$.
\begin{equation*}\label{region2}
S_1=\frac1{12}\left[\log\left(16(2-{x}-{x}^{^{-1}})\right)
+\frac{4({x}-{x}^{^{-1}})}{\pi(2-{x}-{x}^{^{-1}})} 
I\left(\frac1{\sqrt{1-{x}}}\right)I\left(\frac1{\sqrt{1-{x}^{^{-1}}}}\right)
\right].
\end{equation*}
\end{itemize}
Here $I(z)$ is the complete elliptic integral of the first kind
$$I(z)=\int_0^1 \frac{\mathrm{d} y}{\sqrt{(1-y^2)(1-z^2y^2 )}}.$$
As it was discussed before, the fact that the entropy depends solely 
of the parameter $x$ can be derived as a consequence of the M\"obius 
invariance that we uncover in this paper. A consequence of such invariance 
is that the R\'enyi entropy is a function of $x$ as well.
In the following we will use the M\"obius transformations to study some dualities 
and other relations that occur between theories in the different
regions we introduce above.

\subsection{Duality between regions 1a and 1b}

Examining the expressions of the entropy in the region 1, (\ref{region1a}) and (\ref{region1b}), 
it is clear that the entropy is invariant under the change of $x$ to $x^{-1}$. 
We would like to understand this property in the light of the 
symmetries discussed before.

In fact we may derive the duality in the following way. 
Assume we start with a theory with couplings $\gamma_a>0,h_a$ in the
region 1a, i.e., $\gamma_a^2>1-(h_a/2)^2>0$. Therefore, 
the points $z_{a+}$ and $z_{a-}$ are two real zeros inside the unit circle. 
Imagine now that we permute them without permuting their inverses. 
As it was discussed in section \ref{frh}, see also Appendix \ref{det-invariance-app},
the permutation between zeros on the same side of the
unit circle does not affect the entropy, however the cross ratio (\ref{crossratio}) 
is now inverted: $(z_{a-},z_{a+};z_{a+}^{-1},z_{a-}^{-1})=x^{-1}$. Note that from the point of 
view of the corresponding Riemann surface, which in this case is a torus, this permutation of zeros 
is equivalent to cut it along the $a$-cycle, perform a $2\pi$ rotation of one of the borders, and glue them again. 
This is precisely one of the two Dehn twists that generate the modular group $SL(2,\mathbb{Z})$ of the torus.

One may object that the roots in the new pairs $z_{a-},z_{a+}^{-1}$ and
$z_{a+},z_{a-}^{-1}$ are not related by inversion any more, 
as it should be in the XY model. 
Here is where the M\"obius transformations come to the rescue.
By suitably chosing a $SL(2,{\mathbb C})$ transformation 
that does not belong to $SO(1,1)$, 
it is posible to transform the two pairs
above into $z_{b+},z_{b+}^{-1}$ and $z_{b-},z_{b-}^{-1}$ 
with the additional property $\overline z_{b+}=z_{b-}$.
Considering now that the M\"obius transformations leave the entropy 
invariant, we can explain the duality between 
(\ref{region1a}) and (\ref{region1b}). In particular, one may take as M\"obius 
transformation that for which
$$\frac{z_{b+}-z_{b-}}{1+z_{b+}z_{b-}}=-i\frac{z_{a+}-z_{a-}}{1+z_{a+}z_{a-}}.$$

These roots actually correspond to a particular choice of the couplings for the
XY model, $h_b,\gamma_b$, belonging to region 1b, 
which can be related to the original ones by
$$\gamma_b=\sqrt{1-\left(\frac{h_a}{2}\right)^2}, \quad \frac{h_b}{2}=\sqrt{1-\gamma_a^2}.$$
In figure \ref{phase_diagram_xy}, we depict graphically in the plane $(\gamma, h)$ this choice: the point $\triangle$
has coordinates $(\gamma_a, h_a)$ and $\bigtriangledown$, with coordinates
$(\gamma_b, h_b)$, is its dual.

It should be noticed that the duality, that is manifest for von Neumann
also holds for the R\'enyi entropy (\ref{entintegr}), as it is based on the 
equality of the determinants $D_X(\lambda)$
for the two values of the couplings. 
On the other hand, in order to establish the duality
we must perform a M\"obius transformation, which implies
that the dispersion relation has changed as a homogeneous field
of dimension $L=1$. Therefore, in this case the spectrum of the Hamiltonian
transforms non trivialy.

Before proceeding we examine for the XY model how the expressions for a general 
Riemann surface specializes to one with genus 1. In this case the Riemann theta function in $\g$ complex variables
reduces to the elliptic theta function with characteristics in one 
variable $\chartheta{\mu}{\nu}(s\,|\,\tau)$. The period matrix $\Pi$ is replaced by the modulus $\tau$ defined by 
\begin{equation}\label{tau-hg}
 \tau=i\frac{I(\xi)}{I(\sqrt{1-\xi^2})},
\end{equation}
with
$$\xi=
\begin{cases}
\sqrt{x},&\quad0<x<1,\quad {\rm case\ 1a},\\
\sqrt{x^{^{-1}}},&\quad x>1,\quad {\rm case\ 1b},\\
\frac1{\sqrt{1-{x}^{^{-1}}}},&\quad x<0,\quad {\rm case\ 2},
\end{cases}
$$
and the order chosen for the zeros and poles of $g^2(z)$ is
$$z_{a+},z_{a-},z_{a+}^{-1},z_{a-}^{-1},\quad {\rm for\ case\ 1a},$$ 
$$z_{b+},z_{b-},z_{b+}^{-1},z_{b-}^{-1},\quad {\rm for\ case\ 1b},$$ 
$$z_{2+}^{-1},z_{2-},z_{2-}^{-1},z_{2+},\quad {\rm for\ case\ 2},$$
that fulfills in all cases the requirement  
of the previous section, i. e. the first two branch points 
are inside the unit circle and the last two outside. 
Consequently, the assignement of indices
is $(+1,+1,-1,-1)$ for the cases 1a, 1b and  
$(-1,+1,-1,+1)$ for the case 2.
Now, one can easily compute the characteristics to give
$\mu_a=\mu_b=-1/2, \nu_a=\nu_b=0$ for cases 1a, 1b and $\mu_2=0,\nu_2=0$ 
in case 2.

\subsection{ Duality between regions 1a and 2: Kramers-Wannier duality}

We derived above a duality for the entanglement entropy that was based in
the exchange of the two zeros on the same side of the unit circle. 
This does not change the Riemann surface, 
nor the characteristics,
and leaves the entropy invariant. Another duality that preserves
the Riemann surface is the exchange of a real root with its inverse. This 
establishes a relation between the regions 1a and 2. This corresponds to $z_{a+}=(z_{2+})^{-1}$. 
The relation between the coupling constants is then,
$$\frac{h_2}2=\frac2{h_a};\qquad \frac{\gamma_2^2-1}{h_2}=
\frac{\gamma_a^2-1}{h_a}.$$
In figure \ref{phase_diagram_xy}, the point $\blacktriangledown$ in region 2
is the dual of $\triangle$. 

One immediately obtains that under this duality, 
the dispersion relation changes as $$\Lambda_2(\theta)=\frac{2}{h_a}\Lambda_a(\theta),$$ 
which implies that up to a trivial rescaling the spectrum of the Hamiltonian is unchanged.
For the particular value $\gamma_a=1$, in which XY
reduces to the Ising model, the dual theory is also
the Ising model ($\gamma_2=1$) with a different magnetic field. 
In this case this duality coincides with the Kramers-Wannier duality \cite{Kadar} which, as it is notoriously known, is 
very useful for determining the critical point of the Ising model \cite{Kramers1, Kramers2}.

If we now consider the behaviour of the entropy under the duality, 
we observe that the Riemann surface does
not change and henceforth the modulus $\tau$ is invariant.
However, as it was discussed above, the characteristics are transformed
from $\mu_a=-1/2$ to $\mu_2=0$ and therefore $D_X(\lambda)$ and the entropy 
are modified. 

This duality can be generalized to Hamiltonians of higher 
range $L$. We can exchange one real zero of $g^2(z)$ by its inverse, 
which is a pole of $g^2(z)$, or one complex zero of $g^2(z)$ and its complex conjugate
by their inverses, which are poles of $g^2(z)$. Then both the Riemann surface and the 
spectrum of the Hamiltonian remain invariant.
However, the characteristics change
and then entanglement entropy varies too. 
Observe that two regions of the space of couplings
related by one of these dualities are separated by a critical hypersurface.

Note that the two dualities between 1a and 1b regions 
and 1a and 2 are of different nature. In the former the entanglement entropy
is invariant while the spectrum of the Hamiltonian changes. On the contrary,
in the latter the entanglement entropy varies
and the spectrum of the Hamiltonian is preserved.

\subsection{Relation between the dual theories in 1a and 2, with region 1b}

A beautiful result is that although the two dual theories in regions 1a and 2 have different 
entanglement entropies, they can be combined to obtain the entropy of a Hamiltonian in the region 1b. 
This result was first noticed in \cite{Igloi} for the Ising line ($\gamma=1$). 
Our results extend this property to other points in the coupling space. 

In order to proceed, recall the useful identity of theta functions \cite{Igusa} that combines the
two dual theories,
$$\hattheta{\mu+1/2}{\nu}(s\,|\,\tau)\ \hattheta{\mu}{\nu}(s\,|\,\tau)
= \hattheta{\mu+1/2}{\nu}(s\,|\,\tau/2).$$
If we apply this identity for $\mu=0,\nu=0$ to the expression 
(\ref{determinant}) for the determinant we deduce that by summing up
the R\'enyi entanglement entropy for two dual theories, like those 
with coupling constants $\gamma_a,h_a$ and $\gamma_2,h_2$ which (see (\ref{tau-hg}) and
the previous section)  
correspond to the same modulus $\tau$, 
we can obtain the R\'enyi entropy of a third theory in the region $1b$
with coupling constants $\gamma_{_T},h_{_T}$ such that its modulus is $\tau/2$.

In particular, we may take
$$
h_{_T}/2=\sqrt{1-\gamma_a^2}, \qquad 
\gamma_{_T}=\gamma_a\frac{1-\sqrt{1-x_a}}{1+\sqrt{1-x_a}},
$$
with
$$
x_a=\frac{1-(h_a/2)^2}{\gamma_a^2}.
$$
Then if we introduce $y=\sqrt{1-x_a}$, by application of 
(\ref{equx}), (\ref{tau-hg}) we obtain
$$\tau\equiv\tau_{_a}=i\ \frac{I(\sqrt{1-y^2})}{I(y)}\quad {\rm and}\quad
\tau_{_T}=i\ \frac{I\left(\displaystyle\frac{1-y}{1+y}\right)}
{I\left(\displaystyle\frac{2\sqrt{y}}{1+y}\right)},
$$ 
and using the following Landen identities for elliptic functions \cite{Abramowitz},
$$I\left(\displaystyle\frac{2\sqrt{y}}{1+y}\right)=(1+y)I(y),
\quad{\rm and}\quad 
I\left(\displaystyle\frac{1-y}{1+y}\right)=\frac{1+y}2 I(\sqrt{1-y^2}),
$$
we obtain $\tau_{_T}=\tau/2$ as we claimed.

\begin{figure}[H]
  \centering
    \resizebox{15cm}{12cm}{\includegraphics{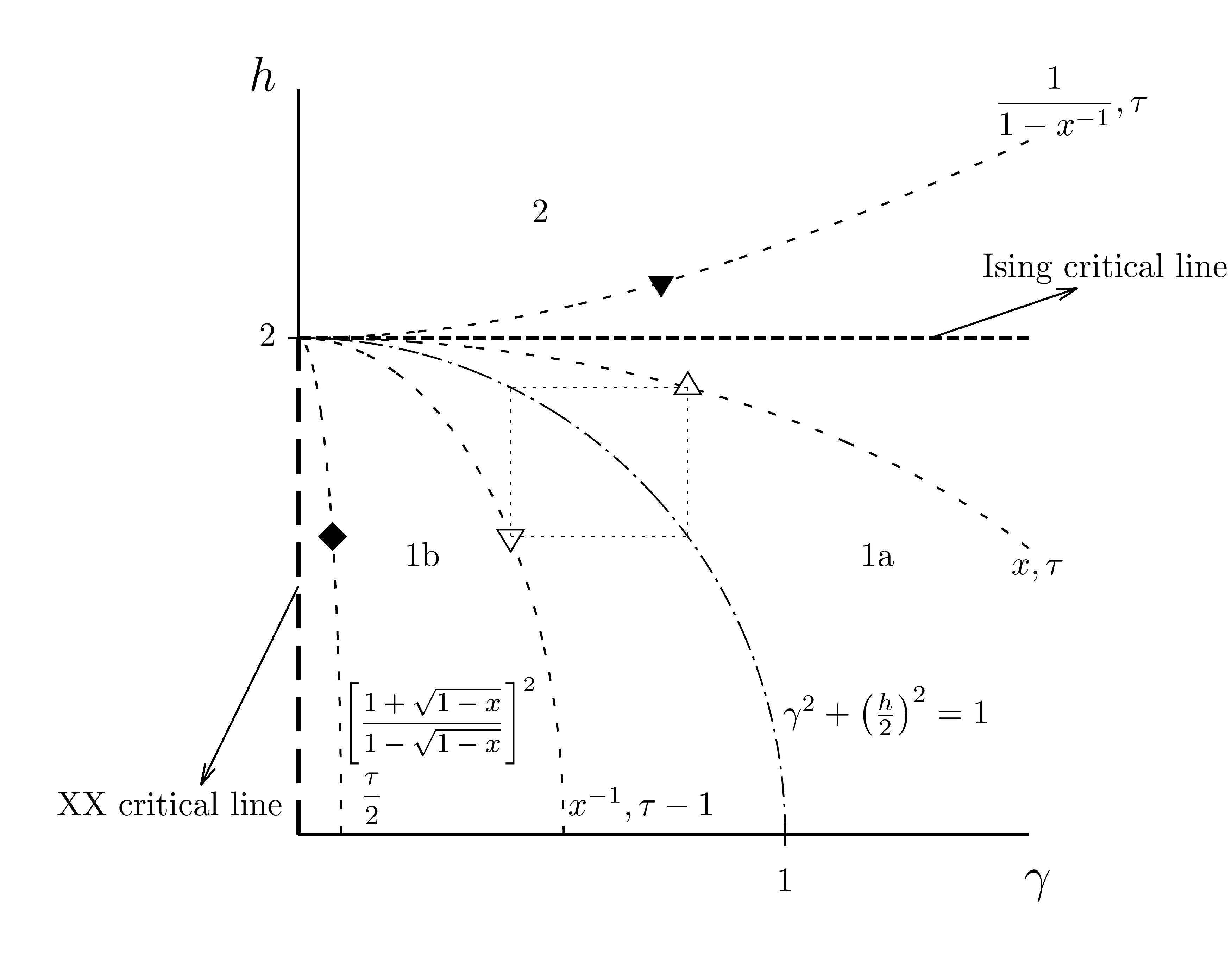}}
    \caption{Plane of couplings $(\gamma, h)$ for the XY Hamiltonian. According to the 
    expression for the entanglement entropy, we distinguish three different regions: 1a, 1b and 2. Dashed
    curves represent the induced flow of $SO(1,1)$. Therefore, each one connects theories with the same
    entanglement entropy. We also depict the different dualities and relations we have studied in the text.
    For each one it is written the cross ratio
    of the branch points and the modulus of the associated torus.
    The dual theory of $\vartriangle$ in 1b is $\triangledown$, $S_\alpha^\triangledown=S_\alpha^\vartriangle$.
    The dual theory of $\vartriangle$ in 2 is $\blacktriangledown$ and, although
    the corresponding tori have the same modulus, $S_\alpha^\blacktriangledown \neq S_\alpha^\vartriangle $. 
    The entanglement entropy of the theory $\blacklozenge$ is given by (\ref{triality}), $S_\alpha^\blacklozenge =
    S_\alpha^\vartriangle +S_\alpha^\blacktriangledown $.}
  \label{phase_diagram_xy}
   \end{figure}

The asymptotic relation of the R\'enyi 
entropies for non critical Hamiltonians reads
\begin{equation}\label{triality}
S_\alpha^a+S_\alpha^2=S_\alpha^T,
\end{equation}
where the superindices obviously refer to 
the theories in the corresponding 
regions with the coupling constants defined above. 
For example, in figure \ref{phase_diagram_xy}, 
the point $\triangle$ is at $(\gamma_a, h_a)$ and $\blacktriangledown$
at  $(\gamma_2, h_2)$; then we have depicted $\blacklozenge$ at $(\gamma_T, h_T)$,
so its entanglement entropy is the sum of entanglement 
entropies of the two other points according
to (\ref{triality}).

The relation is particularly simple when $\gamma_a=\gamma_2=1$ 
(Ising model) in which case the final theory corresponds to the XY model
with zero transverse magnetic field
$h_{_T}=0$ and anisotropy parameter 
$$\gamma_{_T}=\frac{1-h_a/2}{1+h_a/2}.$$
A peculiarity of this case is that the relation (\ref{triality}) for the 
entanglement entropy holds not only in the asymptotic limit but also with
finite size $|X|$, in which case it reads
$$S_\alpha^{a}(|X|)+S_\alpha^{2}(|X|)=S_\alpha^{T}(2|X|).$$
This relation has been known for some time \cite{Igloi} and, in particular,
it has been used to derive the entanglement entropy for the critical 
Ising model using the known results for XX.

\begin{figure}[H]
    \includegraphics[width=1\textwidth]{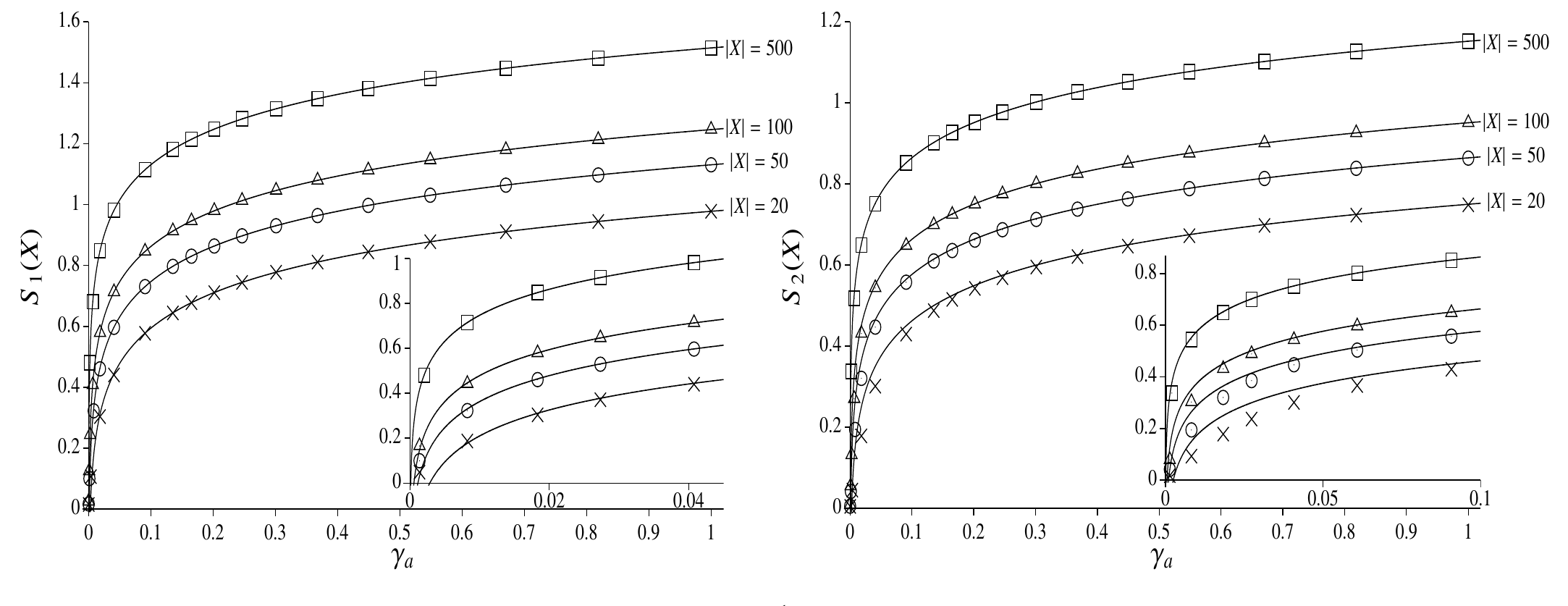}
    \caption{Numerical check of expression (\ref{ising}) for the entanglement
    entropy along the critical Ising line. We have computed the von Neumann 
entropy ($\alpha\to 1$, left panel) and the R\'enyi entanglement entropy
    with $\alpha=2$ (right panel) for different lengths of $X$, varying $\gamma_a$. The insets are a zoom
    of the plot for small values of $\gamma_a$ where the finite size effects 
are more relevant, especially in the case of the R\'enyi entanglement entropy.
}
  \label{ising_line_check}
   \end{figure}

Another simple instance is when we consider the 
critical lines of the Ising and XX universality classes, $h_a=2$ and
$\gamma_b=0$ respectively. In this case, we can establish a duality
between the R\'enyi entropies of the models.
If we take $h_a=2$ and $|\gamma_a|\le1$, the other two related models 
have coupling constants $h_2=2$, $\gamma_2=\gamma_a$, 
$\gamma_{_T}=0$ and $h_{_T}/2=\sqrt{1-\gamma_a^2}$. 
Before writing explicitly the relation of the entanglement entropies
we must consider that we are dealing with critical theories, which implies that
the entropy of an interval $X$ scales logarithmically with its length $|X|$.
As we did for the finite size case, in order to recover the additive relation 
between the entropies in the critical case, we must take different 
length intervals for the different theories, namely  
$$S_\alpha^{T}(2|X|)=2S_\alpha^{a}(|X|).$$
This is an interesting relation because, combined with the results of 
\cite{Jin,AEFS} for the XX model, it allows to compute the R\'enyi 
entanglement entropy for the critical Ising universality class ($h_a=2$).  
The final result is 
\begin{equation}\label{ising}S_\alpha^a(|X|)=\frac{\alpha+1}{12\alpha}\log(|X||\gamma_a|)
+\mathcal{U}_{\mathrm{Ising}}^\alpha, \end{equation}
where 
$$\mathcal{U}_{\mathrm{Ising}}^\alpha=\frac{\alpha+1}{6\alpha}\log 2
+\frac{1}{2\pi i}\int_{-1}^1 \frac{\mathrm{d} f_\alpha (1, \lambda)}{\mathrm{d}\lambda}
\log\frac{\Gamma(1/2-\beta(\lambda))}{\Gamma(1/2+\beta(\lambda))}\mathrm{d}\lambda,$$
with $\Gamma$ the Gamma function. We remark that the expression (\ref{ising}) for the entanglement
entropy of the critical Ising line is a new result. Only the case  $|\gamma_a|=1$ was previously known
in the literature, see \cite{Igloi, Cardy}. 
In Fig. \ref{ising_line_check} we check that this formula matches 
with numeric computations. Note that, since it is an asymptotic result, 
we have finite size effects which are more important
when the branch points approach the unity, i.e. $\gamma_a\to0$.

\section{Conclusion}\label{conclusion-section}

We have seen in this work the importance of casting the space of coupling
constants of a model in terms of a Riemann surface. This simple fact shows up when 
we are computing the asymptotics of the determinant of the $2$-point correlation 
matrix of the model. The next observation is that this determinant is kept invariant 
under M\"obius transformations acting on the Riemann surface that, on the other hand,
change the value of the couplings. Therefore we can have 
changes in the Hamiltonian that leave the determinant invariant. This fact allows us 
to uncover the origin and set up in a unified language the symmetries and dualities of such models. 

Now, the entanglement entropy is a functional of this determinant. Therefore the many suggestions that the 
entanglement entropy is the right tool to study phase transitions find their explicit realization in the 
kind of analysis we perform in this work.

As a next step, we will extend the analysis of the M\"obius transformations 
to the case of critical Hamiltonians \cite{Ares4}. This requires some other 
techniques since, because of the confluence of couples of branch points in the unit circle,
new singularities appear.

There are several directions to generalize our results. One could consider
the case of several disjoint intervals, as mentioned at the end of section \ref{mt-section},
or take complex coupling constants that
break reflection and charge conjugation symmetry \cite{Kadar, Ares3}.
It would be also interesting to apply our analysis to excited states \cite{Alba, AEFS} and to the evolution of the
entanglement entropy after a quantum quench 
\cite{Sengupta, Calabrese, Fagotti2}
in order to determine whether the M\"obius symmetry and the different dualities
hold also in these cases.

\noindent{\bf Acknowledgments:} Research partially supported by grants 2014-E24/2,  
DGIID-DGA and FPA2015-65745-P, MINECO (Spain). FA is supported by FPI Grant No. C070/2014, DGIID-DGA/European Social Fund. 
ARQ is supported by CAPES process number BEX 8713/13-8 and by CNPq under process number 305338/2012-9.

\appendix
\section{Determinant is invariant under permutations of branch points}\label{det-invariance-app}

In this appendix we will check the consistency of the
expression for $\log{D_X(\lambda)}$ that we introduced in (\ref{determinant}).
In particular we will show that it
does not depend on the order in which we choose the roots of
$P(z)$, provided we take the first half inside and the last half outside of the unit circle

Suppose that we exchange the order of  two roots $z_{j_1}$ and $z_{j_2}$, with $j_\kappa=2r_\kappa+1+u_\kappa, \kappa=1,2$ and $r_\kappa=1,\dots,\g$ and $u_\kappa=0,1$.
If we follow the prescriptions of section \ref{frh} this induces a change in the fundamental cycles which are transformed into $a'_r$, $b_r'$, such that
\begin{equation}\label{newcycles}
a_r=
\begin{cases}
a'_r,&r\not=r_1,r_2,\\
a'_{r_1}+\Delta,&r=r_1,\\
a'_{r_2}-\Delta,&r=r_2
\end{cases},\qquad
b_r=
\begin{cases}
b'_r,&r<r_1+u_1,\\
b'_r+\Delta,&r_1+u_1\le r\le r_2-1+u_2,\\
b'_r,&r>r_2-1+u_2
\end{cases}
\end{equation}
where
$$
\Delta=b'_{r_2}-b'_{r_1}+\sum_{t=r_1+u_1}^{r_2-1+u_2}a'_t.
$$
This transformation is a particular instance of the most general 
change of basis of cycles given by a modular transform \cite{Igusa}
$$
\begin{pmatrix}
b\\a
\end{pmatrix}=
\begin{pmatrix}
A&B\\C&D
\end{pmatrix}
\begin{pmatrix}
b'\\a'
\end{pmatrix},\qquad \begin{pmatrix}
A&B\\C&D
\end{pmatrix}\in Sp_{2\g}({\mathbb Z}).
$$

The new period matrix is
$$
\Pi'=(A-\Pi C)^{-1}(\Pi D-B).
$$
While the normalized theta functions are related by
$$
\hattheta{\vec {p'}}{\vec {q'}}(\vec {s'}\,|\,\Pi')
=
{\rm e}^{-\pi i {\vec s}C\cdot{\vec {s'}}}
\hattheta{\vec p}{\vec q}(\vec s\,|\,\Pi)
$$
where 
\begin{equation}\label{newargument}
\vec{s'}=\vec s\,(C\Pi'+D),
\end{equation}
and the characteristics verify
$$({\vec {p'}},{\vec {q'}})
\begin{pmatrix}
b'\\a'
\end{pmatrix}
=
({\vec {p}},{\vec {q}})
\begin{pmatrix}
b\\a
\end{pmatrix}
-\frac12\left({\rm diag}(C^TA),{\rm diag}(D^TB)\right) 
\begin{pmatrix}
b'\\a'
\end{pmatrix}.
$$

In the particular case of the transformation (\ref{newcycles}),
after a straightforward calculation, one obtains
\begin{equation}\label{newcharacteristics}
({\vec {p'}},{\vec {q'}})
\begin{pmatrix}
b'\\a'
\end{pmatrix}
=
({\vec {p}},{\vec {q}})
\begin{pmatrix}
b\\a
\end{pmatrix}
+\frac12(u_1+u_2)(b'_{r_1}+b'_{r_2})
+\frac12(u_1+u_2-2)\sum_{t=r_1+u_1}^{r_2-1+u_2}a'_t.
\end{equation}

We shall examine now how the arguments of the theta 
functions in (\ref{determinant}), $\vec s=\pm\beta(\lambda)\vec e$,
are modified by the transposition.
Taking into account 
the definition of $\vec e$ and the form of the matrices $C$ and $D$ 
one has $\vec e\, C=0$ and $\vec e\, D=\vec e$ if and only if the two roots $z_{j_1}$ and $z_{j_2}$
that we exchange verify $j_1,j_2\le 2L$ or
$j_1,j_2>2L$, which means that both roots sit at the same side of the unit circle.
In this case applying (\ref{newargument}) one has $\vec {s'}=\vec s=\pm\beta(\lambda)\vec e$.

In (\ref{characteristics}) we gave a prescription to obtain the characteristics
for the theta functions involved in the computation of the entanglement 
entropy. They depend on the position occupied by the poles and zeros of $g^2(z)$
which are labeled by a sign $\epsilon_j$. If we exchange two of them, the original 
characteristics $\vec \mu$, $\vec \nu$ change into
\begin{equation}\label{tildecharacterisitics}
\tilde\mu_r=
\begin{cases}
\mu_r,&r\not=r_1,r_2,\\
\mu_{r_1}+\frac14\delta,&r=r_1,\\
\mu_{r_2}-\frac14\delta,&r=r_2,
\end{cases}\qquad
\tilde\nu_r=
\begin{cases}
\nu_r,&r<r_1+u_1,\\
\nu_r+\frac14\delta,&r_1+u_1\le r\le r_2-1+u_2,\\
\nu_r,&r>r_2-1+u_2,
\end{cases}
\end{equation}
where $\delta=\epsilon_{j_2}-\epsilon_{j_1}$. These, in general, 
are different from those obtained by the application of 
(\ref{newcharacteristics}) to $\vec\mu$ and $\vec \nu$ which we denote by $\vec{\mu'}$ and $\vec {\nu'}$.
After a somehow lengthy but direct computation 
one obtains
\begin{eqnarray}
(\vec{\tilde\mu}-\vec{\mu'},
\vec{\tilde\nu}-\vec{\nu'})
\begin{pmatrix}
b'\\a'
\end{pmatrix}
&=&
\left(u_1\frac{\epsilon_{j_1}-1}2
-
u_2\frac{\epsilon_{j_2}+1}2\right)
b'_{r_2}
-
\left(u_1\frac{\epsilon_{j_1}+1}2
-
u_2\frac{\epsilon_{j_2}-1}2\right)b'_{r_1}
\cr&&+
\left((1-u_1)\frac{\epsilon_{j_1}-1}2
-
(1-u_2)\frac{\epsilon_{j_2}+1}2\right)
\sum_{t=r_1+u_1}^{r_2-1+u_2}a'_t.
\end{eqnarray}

The important point to notice here is that,
given that $u_\kappa=0,1$ and $\epsilon_j=\pm1$, one always has
$
\vec{\tilde\mu}-\vec{\mu'},
\vec{\tilde\nu}-\vec{\nu'}\in{\mathbb Z}^\g.
$ 
This implies that
$$
\hattheta{\vec{\tilde\mu}}{\vec{\tilde\nu}}=
\hattheta{\vec{\mu'}}{\vec{\nu'}},
$$
as one can easily check from the definitions.

Finally, putting everything together one has
$$
\hattheta{\vec{\tilde\mu}}{\vec{\tilde\nu}}(\pm\beta(\lambda){\vec e}\,|\,\Pi')=
\hattheta{\vec{\mu'}}{\vec{\nu'}}(\pm\beta(\lambda){\vec e}\,|\,\Pi')=
\hattheta{\vec{\mu}}{\vec{\nu}}(\pm\beta(\lambda){\vec e}\,|\,\Pi),
$$
where for the second equality we assume that the two branch points, whose order was exchanged,
belong both to the first half of the ordering ($j_1,j_2\le 2L$) or both to the second half ($j_1,j_2> 2L$).

From the equality of the normalized theta-functions we deduce
that a change in the order in which we take the roots does 
not affect the expression for the determinant, provided we do not 
exchange a root inside the unit circle with one outside. 

\section{M\"obius transformations on the Space of homogeneous polynomials}\label{vilenkin-app}

In order to obtain how Laurent polynomials (\ref{laurent}) and, accordingly, our 
Hamiltonian (\ref{hamiltonian}) change under M\"obius transformations we can
study the representations of $SL(2, \mathbb{C})$ on the space
of homogeneous polynomials of two complex variables \cite{Vilenkin}. 
In general, adopting the passive point of view,
to each element $V=\left(\begin{array}{cc}a&b\\c&d\end{array}\right)
\in SL(2,{\mathbb C})$
corresponds the linear transformation in $\mathbb{C}^2$
$$(z_1, z_2)\mapsto (d z_1-b z_2, -c z_1+ a z_2).$$
Associated with this transformation we have the operator 
$T_V$ which acts on the space of functions $f: \mathbb{C}^2\to \mathbb{C}$,
such as
$$T_V f(z_1, z_2)=f(d z_1-b z_2, -c z_1+az_2).$$
Note that $T_V$ is a reducible representation of $SL(2, \mathbb{C})$
in the space of functions of two complex variables because this space 
contains an invariant subspace under
$T_V$: the space $\mathfrak{H}_{2L}$ of homogeneous, two variable 
polynomials of degree $2L$,
$$h(z_1, z_2)=\sum_{j=-L}^L u_j z_1^{L+j} z_2^{L-j}.$$

The restriction of $T_V$ to $\mathfrak{H}_{2L}$, $T_V^\mathfrak{H}$,
is an irreducible representation of $SL(2, \mathbb{C})$ on the space
of two complex variable functions. 

We can realize the space of Laurent polynomials $\mathfrak{L}_L$ of degree $L$ in one
(complex) variable from $\mathfrak{H}_{2L}$. In fact, any Laurent polynomial
$$\Upsilon (z)=\sum_{l=-L}^L u_l z^l$$
can be expressed like 
$$\Upsilon (z)=z^{-L} h(z, 1).$$
Therefore, taking into account the homogeinity of $h(z_1, z_2)$, we arrive
at the representation of $SL(2, \mathbb{C})$ in the 
space of Laurent polynomials

\begin{equation}\label{moebius-laurent2}
T_V^\mathfrak{L} \Upsilon (z)=(d z-b)^{L} (-c+ a z^{-1})^{L} 
\Upsilon \left(\frac{d z-b}{-c z+a}\right),
\end{equation}
which is just the passive point of view of expression (\ref{moebius-laurent}).

Now let us choose as a basis of $\mathfrak{L}_L$ the monomials $\{z^m\}$ with
$-L\leq m \leq L$, and the $SU(2)$-invariant scalar product for which  
\begin{equation}\label{scalar-product}
 (z^{m}, z^{n})=\delta_{mn}.
\end{equation}
The matrix elements of $T_V^\mathfrak{L}$ in this basis are
$$t_{mn}^{(V)}=( z^m,T_V^\mathfrak{L} z^n).$$
After a bit of algebra and using (\ref{moebius-laurent2}), 
(\ref{scalar-product}) and Newton's binomial theorem
we arrive to
\begin{multline*}
 t_{mn}^{(V)}=[(L+m)!(L-m)!(L+n)!(L-n)!]^{-1/2}\\
\sum_{j=\max(0, m-n)}^{\min(L-n, L+m)} {L-n \choose j}{L+n \choose L+m-j}
(-1)^{n-m}a^{L-n-j}b^j c^{n-m+j} d^{L+m-j}.
\end{multline*}

Therefore the coefficients of the Laurent polynomial 
$\Upsilon(z)$ transform under $V\in SL(2,\mathbb{C})$ as follows 
$$u'_l=\sum_{m=-L}^L t_{lm}^{(V)}u_m .$$
Since the coefficients of our Laurent polynomials $\myPhi(z)$ and $\Xi(z)$
are precisely the couplings of our Hamiltonian (\ref{hamiltonian}), we have just 
found their behaviour under M\"obius transformations,
$$A'_l=\sum_{m=-L}^L  t_{lm}^{(V)}A_m;\quad B'_l=\sum_{m=-L}^L t_{lm}^{(V)}B_m.  $$

Note that $t_{m,n}^{(V)}=t_{-m,-n}^{(V)}$ when $a=d$ and $b=c$, as it happens for 
$SO(1,1)$. Therefore, in that case, the new couplings
satisfy the requiered properties $A_{-l}=A_l$ and $B_{-l}=-B_l$.

  \end{document}